\documentclass[prb,aps,twocolumn,showpacs,preprintnumbers,amsmath,amssymb]{revtex4}
\usepackage{dcolumn}
\usepackage{graphicx}
\usepackage{epstopdf}

\begin{document}

\title{Density functional calculations of the electronic structure and
magnetic properties of the hydrocarbon K$_{3}$picene superconductor near
the metal-insulator transition}

\author{Minjae Kim and B. I. Min$^{*}$}
\affiliation{Department of Physics, PCTP,
	Pohang University of Science and Technology, Pohang, 790-784, Korea}
\author{Geunsik Lee, Hee Jae Kwon, Y. M. Rhee and Ji Hoon Shim$^{*}$}
\affiliation{Department of Chemistry,
    Pohang University of Science and Technology, Pohang, 790-784, Korea}
\date{\today}

\begin{abstract}
We have investigated the electronic structures and magnetic properties
of a newly discovered hydrocarbon superconductor, K$_{3}$picene,
having $T_{c}$=18 K.
We have shown that the metal-insulator transition (MIT) is
driven in K$_{3}$picene by 5\% volume enhancement
with the formation of the local magnetic moment.
Active bands for the superconductivity near the Fermi level ($E_F$)
have hybridized character of LUMO and LUMO+1 of the picene molecule.
Fermi surfaces of K$_{3}$picene manifest neither prominent nesting feature
nor marked two-dimensional behavior.
By estimating the ratio of the Coulomb interaction $U$
and the bandwidth $W$ of the active bands near $E_F$,
we have demonstrated that K$_{3}$picene is located in the vicinity of
the Mott transition.
Our findings suggest that K$_{3}$picene is a strongly correlated electron
system.
\end{abstract}


\maketitle

\begin{figure}[b]
\includegraphics[width=9cm]{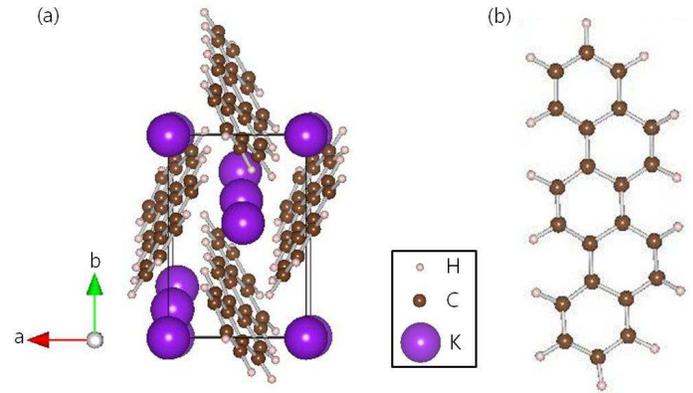}
\caption{(Color online)
(a) Crystal structure of K$_{3}$picene.
There are two stacked picene molecules in a unit cell.
K atoms are intercalated between the
stacked picene molecules.
(b) Molecular structure of a picene.
}
\label{fig1}
\end{figure}

Recently, the first hydrocarbon superconductivity has been observed in the K-doped picene,
K$_{3}$picene\cite{mitsuhashi}.
Its transition temperature $T_{c}$=18 K
is comparable to that of the alkali-metal doped fullerene system
($T_{c}\sim$38K)\cite{takabayashi}.
In K$_{3}$picene, K atoms are intercalated in the stacked picene
molecules, as shown in Fig.~\ref{fig1}(a).
A picene molecule (C$_{22}$H$_{14}$) consists of five connected benzene
rings with armchair edge (see Fig.~\ref{fig1}(b)).
Interestingly, alkali metal-doped $A_{3}$picenes
($A$=Na, K, Rb, and Cs) exhibit the cation $A$-dependent physical
properties\cite{mitsuhashi}.
With increasing the cation size, the system varies from
a Pauli-like paramagnet for $A$=Na to superconductors for $A$=K and Rb,
and then to an insulator for $A$=Cs.
This feature suggests that $A_{3}$picene is susceptible to
the chemical pressure.
This trend is also reminiscent of $A_{3}$C$_{60}$ ($A$=K, Rb, and Cs),
in which changing cation from K to Cs results in 6\% volume
expansion so as to induce the metal-insulator transition (MIT)\cite{huang}.

Unconventional superconductors
often manifest the incipient MIT and magnetic instability,
which reflect the existence of strongly correlated
electrons\cite{lee,capone}.
The parent compounds of cuprate high $T_c$ superconductors are typically
antiferromagnetic (AFM) insulators\cite{aharony}.
Also the organic superconductors $\kappa$-(BEDT-TTF)$_{2}$X and
$A_{3}$C$_{60}$ ($A$=K and Cs) show the phase transition from an AFM insulator
to a superconductor upon increasing pressure\cite{lefebvre,takabayashi}.
These examples suggest that investigation of the electronic structures
of K$_{3}$picene under volume change would provide an important clue to
the pairing mechanism of superconductivity.

In this report, we have explored the volume-dependent electronic and magnetic
properties of K$_{3}$picene using the first principles band structure
calculation within the density functional theory.
We have shown that the
MIT and AFM transitions are driven in K$_{3}$picene
by $5$\% volume enhancement
with the formation of the local magnetic moment.
Our results indicate that K$_{3}$picene itself is on the verge of
both MIT and magnetic instability.
We have found that the active bands for superconductivity
have hybridized character of LUMO and LUMO+1 of the picene molecule.
Quantitative estimation of $U/W$ ($U$:Coulomb interaction, $W$: band width)
also provides that K$_{3}$picene is near to the Mott transition.


\begin{figure}[t]
\includegraphics[width=8cm]{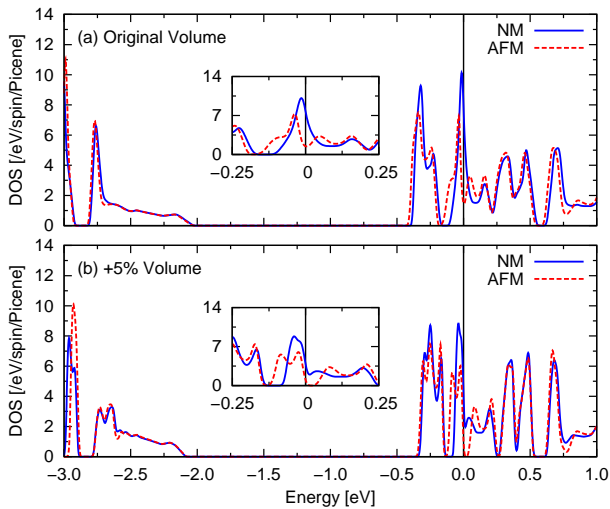}
\caption{(Color online)
DOS for NM (blue solid lines)
and AFM (red dotted lines) phases of K$_{3}$picene.
Inset shows the amplified DOS near $E_F$.
(a) K$_{3}$picene with  the original volume (K$_{3}$P[OV]),
and (b) K$_{3}$picene with 5\% volume expansion (K$_{3}$P[EV]).}
\label{fig2}
\end{figure}

In the unit cell of K$_{3}$picene,
there are two stacked picene molecules and six K atoms (see Fig.~\ref{fig1}(a)).
The positions of K atoms are still undetermined experimentally.
Hence, we first optimized the internal atomic positions in K$_{3}$picene
with retaining the observed monoclinic structure
(lattice constants: a=5.912, b=8.707,
c=12.97 ${\AA}$, and $\alpha$=92.77$^{\circ}$).\cite{mitsuhashi}
For this purpose, we have used the pseudo-potential band method of
VASP package\cite{VASP}.
For the exchange-correlation potential, the generalized gradient
approximation (GGA) of Perdew-Burke-Ernzerhof
(PBE) was utilized\cite{GGA}.
We have used 100 {\bf k}-points inside the first Brillouin zone.
The convergence of the total energy
with respect to number of {\bf k}-points was checked to have
precision of less than 4 meV per formula unit.
For the structural optimization, the atomic relaxations are terminated
when the forces in all atomic sites are less than 0.05 eV/{\AA}.
The stable positions of K atoms are
determined as four above the end benzene rings and
two above the center rings with nearly two dimensional
stacked arrangement of picene molecules\cite{OS},
similarly to the results of Ref. \cite{kosugi,andres}.
However, the arrangement of picene molecules in Ref. \cite{kosugi}
are perpendicular to $ab$ plane losing the
conventional herringbone structure of the pristine picene solid.
Our optimized structure retains the conventional herringbone
structure in accordance with Ref. \cite{andres}.
We have also optimized the internal atomic positions of K$_{3}$picene
with 5\% volume expansion in order to examine the chemical pressure
effect, as is realized in Cs$_{3}$picene.
Then, based on the optimized structures,
we have obtained the electronic structure of K$_{3}$picene by
employing the more precise
all electron full-potential linearized augmented plane wave (FLAPW)
band method\cite{FLAPW} implemented in WIEN2k package\cite{Blaha}.
To explore the magnetic properties of K$_{3}$picene,
we have considered the AFM spin configuration with
the opposite spin polarizations of two independent picene molecules
in the unit cell.


\begin{figure}[t]
\includegraphics[width=8.3cm]{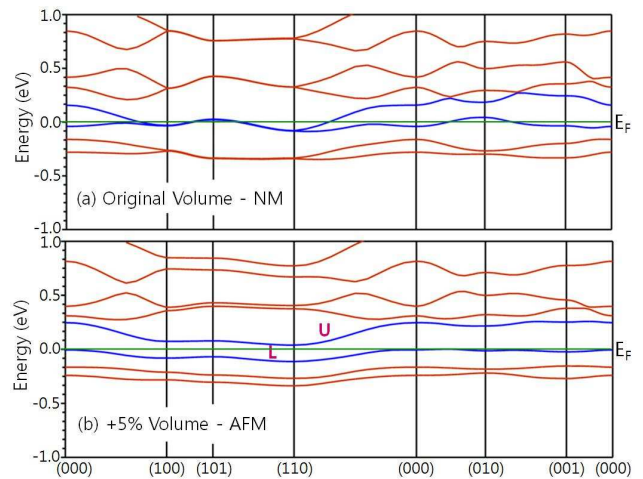}
\caption{(Color online)
(a) Band structure for NM phase of K$_{3}$picene
with original volume (K$_{3}$P[OV]).
(b) Band structure for AFM phase of K$_{3}$picene
with 5\% volume expansion (K$_{3}$P[EV]).
Notice that two bands near $E_F$ are split into U and L which are
represented by blue (dark) color.
}
\label{fig3}
\end{figure}

Figure~\ref{fig2} shows the density of states (DOS) of K$_{3}$picene.
DOSs of K$_{3}$picene with original volume (hereafter K$_{3}$P[OV])
and with +5\% volume expansion (hereafter K$_{3}$P[EV])
are provided in their nonmagnetic (NM) and AFM phases.
Pristine solid picene is a semiconductor with a wide band gap
of 3.3 eV.\cite{okamoto}
Figure~\ref{fig2}(a) shows that
K$_{3}$P[OV] also has an energy gap structure
($\sim 1.7$ eV) between HOMO and LUMO related bands.
In K$_{3}$picene, three electrons are transferred
from three intercalated K atoms
into one picene, occupying LUMO and LUMO+1 related bands,
as was pointed out by Kosugi {\it et. al.}\cite{kosugi}.
For K$_{3}$P[EV] in Fig.~\ref{fig2}(b),
the band widths become narrower to have a larger energy gap
($\sim 1.9$ eV) between HOMO and LUMO related bands.
It is seen that, for both K$_{3}$P[OV] and K$_{3}$P[EV],
the DOS at the Fermi level ($E_F$) is much reduced
in the AFM phase with respect to that of the NM phase.
Noteworthy is that K$_{3}$P[EV] becomes an insulator in the AFM phase,
having a small energy gap of $\sim 0.03$ eV.
The NM and AFM phases are almost degenerate.
Nevertheless, the magnetic moment of one picene, $\sim$0.35 $\mu_{B}$,
is large enough to stabilize
the magnetic ground state.\cite{NMvsAFM,giovannetti}
These features suggest that K$_{3}$picene (K$_{3}$P[OV]) is
on the verge of the MIT and magnetic instability.
The AFM structure in K$_{3}$P[EV] with the localized
magnetic moment of a picene
provides the explanation of the significant enhancement of $M/H$
observed in Cs$_{3}$picene\cite{mitsuhashi}.
Like other molecular solids (KO$_{2}$, K-doped pentacene, Cs$_{3}$C$_{60}$)
having local moment carrying 2$p$ electrons\cite{takabayashi,mori,labhart},
the local moment in K$_{3}$P[EV] arises from the unpaired electron
in the localized molecular orbital.

\begin{figure}[t]
\includegraphics[width=8.8cm]{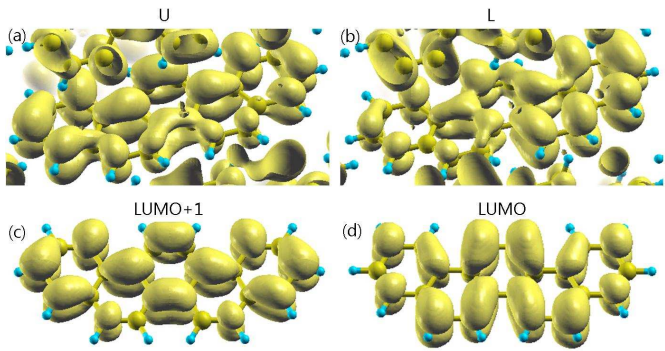}
\caption{(Color online) Charge densities (CDs) of (a) U and (b) L bands
in Fig.~\ref{fig3}.
CDs of (c) LUMO+1 and (d) LUMO orbitals of a picene molecule.
CD of the U band shows mostly LUMO+1 character
with small hybridization with LUMO,
while CD of the L band shows hybridized character of LUMO and LUMO+1.
}
\label{fig4}
\end{figure}

In Fig.~\ref{fig3}(a) and (b), the NM band structure of K$_{3}$P[OV]
and the AFM band structure of K$_{3}$P[EV] are plotted, respectively.
The former exhibits a metallic state with two bands crossing $E_F$,
while the latter shows an insulating state.
The metallic band structure of K$_{3}$P[OV]
results from the almost degenerate U and L bands near $E_F$, while
the insulating state in the AFM phase of K$_{3}$P[EV]
emerges due to the shift up and down of respective U and L bands
that had become narrowed due to the volume expansion.
Notice that the degenerate bands in the NM phase of K$_{3}$P[OV]
at {\bf k}-points along the (100)-(101)-(110) directions
are split in the AFM phase of K$_{3}$P[EV].
The overlap of $\pi$-$\pi$ orbitals of adjacent molecules
along the (100) direction produces the symmetric hopping
along this direction, resulting in the degeneracy on the
(100) zone surface for the NM phases\cite{hotta}.
This degeneracy on the (100) zone surface is lifted
by the AFM spin ordering for both K$_{3}$P[OV] and K$_{3}$P[EV],
due to the broken symmetry along the (100) direction.
This feature is reminiscent of the Slater-type MIT mechanism
accompanied by the cell doubling, and might suggest
that K$_{3}$P[EV] can be categorized into a band insulator.
However, similar feature was observed in the
typical Mott insulating systems such as cuprate and nickel oxides.
Therefore, in order to confirm that
whether K$_{3}$P[EV] is a band or Mott insulator,
one needs to check the $U/W$ value of a system,
as will be discussed later.

We have examined the orbital characters of U and L bands
near $E_F$ that are responsible for superconductivity.
The charge densities (CDs) of U and L bands are plotted in
Fig.~\ref{fig4}(a) and (b), respectively.
For comparison, the CDs of LUMO+1 and LUMO of a picene molecule are
also provided in Fig.~\ref{fig4}(c) and (d), respectively.
The CDs in Fig.~\ref{fig4} reveal that U and L bands
have the mixed character of both LUMO and LUMO+1,
even though LUMO+1 character is dominant for U.
The energy splitting between LUMO and LUMO+1
of a picene molecule is known to be 0.15 eV.\cite{mitsuhashi}
This gap size is comparable to the band widths
of four conduction bands near $E_F$
(U, L, and two lower bands below L in Fig.~\ref{fig3}(b)),
which correspond to 0.21, 0.12, 0.12, and 0.13 eV, respectively.
The comparable size of the gap and band widths
indicates that the hybridization interaction between LUMO and LUMO+1
is large enough to yield the hybridized bands with mixed
LUMO and LUMO+1 character near $E_F$.
Thus, the active bands for superconductivity near $E_F$
have hybridized character of LUMO and LUMO+1 molecular orbitals,
in agreement with the earlier report by Kosugi {\it et. al.}\cite{kosugi}.
This is contrary to the case of pentacene,
in which the energy splitting between LUMO and LUMO+1 ($\sim 1.28$ eV)
is much larger than that of picene.\cite{kosugi}

\begin{figure}[b]
\includegraphics[width=8.8cm]{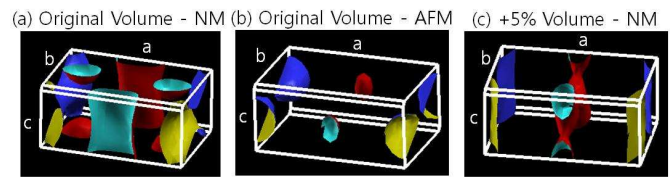}
\caption{(Color online)
Fermi surfaces of K$_{3}$picene.
(a) NM phase of K$_{3}$P[OV],
(b) AFM phase of K$_{3}$P[OV], and
(c) NM phase of K$_{3}$P[EV].
}
\label{fig5}
\end{figure}

Fermi surfaces of K$_{3}$picene are provided in Fig. \ref{fig5}.
Note first that there is no prominent nesting feature
in the Fermi surfaces of K$_{3}$P[OV] (Fig. \ref{fig5}(a)),
which is in agreement with Ref. \cite{kosugi}.
Secondly, there is no Fermi surfaces of clear cylindrical shape
along the $c$-axis, indicating that
this system would not have marked two dimensional behavior,
despite its layered-like crystal structure.
This feature is indeed consistent with the band structure in Fig. \ref{fig3}(a),
which shows apparent dispersions along the $c$-axis,
(100)-(101) and (001)-(000).
Figure~\ref{fig5}(b) and (c) illustrate explicit effects
of AFM spin polarization and volume enhancement on the Fermi surfaces
of K$_{3}$picene, respectively.
For the AFM phase of K$_{3}$P[OV] (Fig. \ref{fig5}(b)),
the volume of Fermi surfaces is reduced much,
with respect to that of the NM phase of K$_{3}$P[OV]
(Fig. \ref{fig5}(a)). It is because of the reduced DOS
near $E_F$ accompanied by the AFM spin polarization.
The volume enhancement also reduces the Fermi surface area,
as shown in Fig. \ref{fig5}(c), due to the decreased band width.

We have calculated electrical conductivity $\sigma$
using the Boltzmann transport theory\cite{allen,madsen}.
The $\sigma/\tau$ ($\tau$: scattering time) value of the NM phase
of K$_{3}$P[OV] is estimated to be about 0.75 ($10^{19}S/m\cdot sec$).
The calculated anisotropy is only about two, which implies that this system is
not so anisotropic, as addressed in Fig. \ref{fig5}(a).
The reduced Fermi surfaces due to the explicit effects of
AFM spin polarization and volume enhancement result in significant
drop in the conductivity.
The $\sigma/\tau$ values of the AFM phase of K$_{3}$P[OV]
and the NM phase of K$_{3}$P[EV] turn out to be 0.25 and
0.13 ($10^{19}S/m\cdot sec$), respectively, which are only 33\% and 17\%
of that of the NM phase of K$_{3}$P[OV] (0.75$\times10^{19}S/m\cdot sec$).
The conductivity analysis indicates that both volume enhancement and AFM spin
polarization are effective for the MIT.
The volume enhancement leads to the heavy effective mass in conductivity
by reducing $W$, while the AFM spin polarization leads to
the reduced DOS at $E_F$ by splitting two bands around $E_F$ farther.

In general, the volume enhancement of solid
decreases the band width $W$ and increases the Coulomb repulsion $U$,
resulting in larger $U/W$ value.
The increased $U/W$, when it becomes larger than the critical value
$(U/W)_{c}$, would drive the MIT through a Mott transition.
Indeed, in K$_{1}$pentacene, the formation of insulating state
is experimentally realized\cite{craciun}.
We have estimated $U/W$ of K$_{3}$picene,
using the method of Ref. \cite{brocks}
that was applied to the pentacene solid.
The effective Coulomb interaction ($U_{eff}$)
is given by subtracting the polarization energy ($E_{pol}$)
from the bare Coulomb interaction ($U_{bare}$),
$U_{eff}=U_{bare}-E_{pol}$.
$U_{bare}$ of a charged picene molecule is estimated
from the second order M{\o}ller-Plesset calculations with
6-31G* atomic orbital basis set for an isolated picene
by considering
different anionic electron occupations\cite{MP20,MP23}.
$E_{pol}$ for a solid K$_{3}$picene can be
determined by using the partial continuum model,
in which a charged molecule
is assumed to be surrounded
by a cavity of homogeneous dielectric medium\cite{brocks}.
We have obtained the dielectric constant and
the size of the cavity by fitting the calculated $E_{pol}$
to the experimental value of the pristine solid picene\cite{sato}.
In this way, we have got $U_{bare}=2.25$ eV and $E_{pol}=1.40$ eV,
respectively, and thus $U_{eff}$ is estimated to be 0.85 eV.
One can estimate the $U/W$ for K$_{3}$P[EV]
by adopting $W$ of L+U bands near $E_F$ (0.30 eV)
for the NM phase of  K$_{3}$P[EV].
Then we have obtained $U/W=2.83$, which is larger than
$(U/W)_{c}=1.73$\cite{UWc,Gunnarsson,roth}, so that
a Mott insulating state is realized in K$_{3}$P[EV].
This large $U/W$ value for K$_{3}$P[EV] suggests that
K$_{3}$P[OV], which exhibits superconductivity,
would also be near the boundary of the Mott transition.

In a recent report on the K atom adsorbed picene molecule\cite{baker},
it was argued that the charge transfer from K atoms
to LUMO and LUMO+1 is not decisive due to the strong
Coulomb correlation effect in picene.
The above analysis for the Coulomb interaction, however, shows that
$U_{eff}$ in K$_{3}$picene solid is only about 38\% of $U_{bare}$
in an isolated picene molecule.
Therefore, distinctly from the case of the picene molecule,
the charge transfer from K atoms to LUMO and LUMO+1 related bands
will be possible in solid picene due to
the reduced effective Coulomb interaction.


In conclusion, we have studied the electronic structures and magnetism
of a first hydrocarbon superconductor, K$_{3}$picene,
by employing the first principles band structure method
within the density functional theory.
Active bands for superconductivity are found to have LUMO and
LUMO+1 hybridized character.
Fermi surfaces of K$_{3}$picene manifest neither prominent nesting feature
nor marked two-dimensional behavior.
The latter is also supported by the rather small anisotropy in the
estimated conductivity.
We have demonstrated that
K$_{3}$picene is on the verge of the MIT and the magnetic instability.
Therefore, one needs to consider a 3D model with strong Coulomb correlation
effects to investigate the mechanism of superconductivity in K$_{3}$picene.

After submission of the manuscript, we became aware
of a related work by G. Giovannetti and M. Capone, Ref.\cite{giovannetti},
in which the AFM state is found to become stabilized in K$_{3}$P[OV]
by the inclusion of correlation effects.
Also recent experiment confirms that the Coulomb interaction
($U$) in solid picene is about 0.85 eV, which is much larger than the
band width, Ref.\cite{roth1}.
These results are consistent
with the present results suggesting that K$_{3}$picene
is a strongly correlated electron system.

\begin{acknowledgments}
The authors acknowledge useful discussion with K. S. Kim and K. Lee.
This work was supported by the NRF (No. 2009-0079947, 2010-0006484),
WCU through KOSEF (No. R32-2008-000-10180-0),
and by the POSTECH BK21 Physics Division.
\end{acknowledgments}

$^*$ bimin@postech.ac.kr; jhshim@postech.ac.kr


\begin{thebibliography}{99}
\bibitem{OS} We have checked that the other possible K$_{3}$picene
	structure, in which K atoms are out of the stacked planar
	picene molecules, is unstable by more than 1 eV per picene,
	as compared to the structure considered in the present study.
\bibitem{NMvsAFM}
     We have used the FLAPW band method for the total energy calculation.
     For K$_{3}$P[EV], the AFM state is more stable
	 than the NM state by 1 meV per formula unit.
     Thus, in K$_{3}$P[EV], the NM and AFM states are nearly degenerate,
     in view of that the total energy difference is less than
     the precision of total energy calculation ($\sim 4$ meV).
     On the other hand, for K$_{3}$P[OV],
     the NM state is more stable than the AFM state
     by 5 meV per formula unit with similar magnetic moment of
     $\sim$0.35 $\mu_{B}$ per one picene.
     The metastable AFM state in K$_{3}$P[OV], which has comparable
     total energy with the NM state,
     was also reported in Ref.\cite{giovannetti}
     by using the pesudo-potential VASP band method.
\bibitem{UWc}
On the basis of (i) strong hybridized character of LUMO and LUMO+1
for L and U bands near $E_F$, (ii) electron energy loss
scattering experiment on pristine solid picene\cite{roth},
and (iii) our second order M{\o}ller-Plesset calculations,
three orbitals (LUMO, LUMO+1, and LUMO+2)
are considered to be nearly degenerate.
Then by assuming the three-fold degeneracy (LUMO, LUMO+1, and LUMO+2),
$(U/W)_{c}$ for the MIT is given by $(U/W)_{c}=\sqrt{3}=1.73$\cite{Gunnarsson}.
\end{thebibliography}
\end{document}